\documentclass[10pt,conference]{IEEEtran}
\IEEEoverridecommandlockouts

\usepackage{cite}
\usepackage{amsmath,amssymb}
\usepackage{graphicx}
\usepackage{booktabs}
\usepackage{array}
\usepackage{multirow}
\usepackage{adjustbox}
\usepackage[table]{xcolor}
\definecolor{grouprow}{gray}{0.90}
\usepackage{tcolorbox}
\usepackage[hidelinks]{hyperref}
\newcommand{\rqpart}[1]{\par\smallskip\noindent\underline{\textit{#1}}.~}
\newcommand{\rqresults}{\par\smallskip\noindent\underline{\textit{Results}}.\par\noindent}
\graphicspath{{figures/}}

\begin{document}

\title{Do AI Coding Agents Write Infrastructure as Code\\ Like Humans? An Empirical Study\\ of GitHub Pull Requests}

\author{
\IEEEauthorblockN{Amine Barrak}
\IEEEauthorblockA{Oakland University\\
aminebarrak@oakland.edu}
\and
\IEEEauthorblockN{Emna Ksontini}
\IEEEauthorblockA{University of North Carolina Wilmington\\
ksontinie@uncw.edu}
\and
\IEEEauthorblockN{Charles Wehbe}
\IEEEauthorblockA{Oakland University\\
cwehbe@oakland.edu}
}

\maketitle

\begin{abstract}
Infrastructure-as-Code (IaC) configurations define cloud resources,
networking, and access control, and AI coding agents increasingly
write them. Little is known about how agent contributions differ from
those of humans and automation bots, how maintainers evaluate them,
and whether they introduce different sustainability concerns.

We study 349,156 IaC pull requests (PRs) from 49,241 GitHub
repositories, distinguishing humans, automation bots, silent agents,
and disclosed agents. Bots account for 32.9\% of ostensibly human PRs,
and 12.3\% of the remaining PRs contain agent traces, a share that grows
over the study period. Agent PRs are about four times larger than human
PRs and more often modify core infrastructure modules, whereas bots
mainly maintain dependencies. Silent-agent PRs have higher overall
merge rates than human PRs (87.4\% vs.\ 78.9\%) but are merged 4.0
percentage points less often within the same repositories. Disclosed
agents create new infrastructure more often than humans (34.0\% vs.\
21.4\%), and even for the same change type within the same repository,
their PRs are 8.4 points less likely to be merged; silent-agent PRs
show comparable acceptance for most change types. Finally, in 2,645
matched agent--human Terraform PR pairs, agent and human PRs introduce
sustainability smells at similar rates (10.4\% vs.\ 10.3\%; paired odds
ratio 1.02, 95\% CI 0.83--1.25), most commonly missing remote state and
monolithic configurations.
\end{abstract}

\begin{IEEEkeywords}
Infrastructure as Code, AI coding agents, automation bots, authorship,
pull requests, empirical study
\end{IEEEkeywords}

\section{Introduction}
\label{sec:intro}

Deploying software requires dependable infrastructure: compute,
networks, storage, identities and access policies, scaling rules, and
persistent state. Infrastructure-as-Code (IaC) specifies these
resources as versioned configurations, so that tools such as Terraform,
Ansible, and CloudFormation can provision them automatically and
repeatably as part of continuous integration and
delivery~\cite{rahman2019mapping,nist800204c,pahl2026aiiac}.
Unlike an ordinary code edit, an IaC change can open a network path,
grant a permission, or provision costly resources once it is
applied~\cite{rahman2019sins,verdet2025policies}. Such changes reach
a project through pull requests (PRs), so each IaC PR is both a
proposed change to the infrastructure and a review decision about
whether to accept it.

Automating deployment does not remove the consequences of configuration
choices. A configuration fixes the characteristics of the resources it
provisions, and these may keep consuming resources long after the
change: an oversized instance type, a fleet of fixed size without
autoscaling, logs retained indefinitely, Terraform state kept on a
local disk, or all resources of a system in a single file. Kosbar and
Hamdaqa~\cite{kosbar2025smellssus} catalogue such decisions as
sustainability smells in IaC and show that whether a smell is wasteful
depends on context. A detected smell is therefore an indicator in the
configuration, not a measurement of energy, cost, or waste.

The growing adoption of AI coding agents introduces a new dimension
to infrastructure automation: beyond automatically provisioning
resources, developers can use AI agents to generate and modify
the configurations that govern those resources. AI coding agents
such as Claude Code, GitHub Copilot, OpenAI Codex, Cursor, and Devin
participate in IaC development, while research
has begun exploring their integration into the IaC and DevSecOps
lifecycle~\cite{pahl2026aiiac,esposito2026copilot}.
Some contributions are explicitly attributed to agents, whereas
others are submitted under developer identities, leaving agent
involvement visible through commit-level
traces~\cite{aidev,census}. Traditional automation bots, such as
Dependabot and Renovate, further complicate attribution because
their contributions are often counted as human
activity~\cite{dey2020bots}. Consequently, identifying who produces
infrastructure changes requires distinguishing humans, traditional
bots, and both disclosed and silent AI agents.

Despite growing interest in AI coding agents, their implications
for real-world IaC development remain insufficiently understood.
Identifying who contributes is only the first step: we also need
to understand which infrastructure operations agents perform,
how maintainers evaluate their changes, and whether these changes
introduce potential resource-efficiency concerns. Existing research
has examined the acceptance and rejection of agent-generated
PRs~\cite{watanabe2025agentic,ehsani2026failed,
nakashima2026rejected}, evaluated the correctness and security of
LLM-generated IaC through benchmarks~\cite{kon2024iaceval,
nekrasov2025iacgen,vargas2026secureterraform}, and investigated
IaC smells without distinguishing human and AI-agent
contributions~\cite{rahman2019sins,rahman2020gang,
war2025sins,kosbar2025smellssus}. However, these studies examine
separate aspects of infrastructure development. In particular,
it remains unclear whether agents perform the same kinds of
infrastructure changes as human developers, whether differences
in their acceptance persist when accounting for change types
and repository context, and whether agent-generated changes
introduce different sustainability-smell patterns. Addressing
these questions requires connecting contributor provenance,
change semantics, review outcomes, and sustainability within
real-world IaC repositories.

We address this gap through a large-scale observational study
of 349,975 IaC pull requests from 49,241 GitHub repositories
opened between January 2025 and May 2026. Our main
contributions are:

\begin{itemize}
    \item \textbf{Authorship and participation.}
    We distinguish four contributor populations across 349,156
    labeled PRs. Automation bots account for 35.8\% of
    ostensibly human PRs, while silent-agent PRs, despite
    higher overall acceptance, are merged 4.0 percentage
    points less often than human PRs within repositories.

    \item \textbf{Infrastructure changes and acceptance.}
    Using an IaC-specific taxonomy and human-audited
    annotations, we find that disclosed agents create new
    infrastructure more frequently than humans
    (34.0\% vs.\ 21.4\%). Their within-repository acceptance
    deficit persists after adjusting for change type
    (8.4 percentage points).

    \item \textbf{Sustainability implications.}
    Across 2,645 matched agent--human PR pairs modifying
    AWS Terraform configurations, we find no statistically
    detectable difference in sustainability-smell introduction
    (10.4\% vs.\ 10.3\%; paired odds ratio 1.02,
    95\% CI 0.83--1.25).
\end{itemize}

\textbf{Replication Package.}
Our data, codebook, prompts, and analysis scripts are
available in the replication package~\cite{replication}.

\section{Related Work}
\label{sec:related}

Our work builds on two lines of research: AI coding agents and
automation in software repositories, and the quality of
Infrastructure-as-Code (IaC) configurations.

\subsection{AI Coding Agents and Automation in Software Repositories}

Early research on AI programming assistants examined the
correctness and security of generated code.
Nguyen and Nadi~\cite{nguyen2022copilot} evaluated GitHub Copilot's
suggestions, while Pearce et al.~\cite{pearce2022asleep}
identified security weaknesses in generated code.
More recently, research has shifted toward autonomous coding agents
and their contributions to open-source projects.
AIDev~\cite{aidev} identifies agent-generated PRs through agent
accounts and signatures, primarily capturing disclosed contributions.
Studies of these PRs have investigated acceptance and rejection:
Watanabe et al.~\cite{watanabe2025agentic} reported lower acceptance
of Claude Code PRs than matched human PRs, often due to project
context; Ehsani et al.~\cite{ehsani2026failed} found that failed
agent PRs tend to be larger and encounter more CI failures;
and Nakashima et al.~\cite{nakashima2026rejected} identified
rejection reasons specific to agent-generated contributions.

However, agent involvement is not always disclosed.
Khosravani and Mockus~\cite{census} demonstrated that commit
trailers, generation footers, and agent identities reveal
contributions missed by PR metadata.
Traditional automation bots complicate contributor
identification. Wessel et al.~\cite{wessel2018bots} characterized
bots in pull-based development, Dey et al.~\cite{dey2020bots}
showed how bot activity affects repository analyses, and
Alfadel et al.~\cite{alfadel2021dependabot} examined developer
responses to Dependabot security PRs.

Our study combines agent-signature detection, commit-level
provenance analysis, and bot identification to distinguish
four contributor populations in IaC repositories. Beyond
characterizing their participation, we investigate the
infrastructure changes they produce and how acceptance
varies across repositories and change types.

\subsection{IaC Quality and AI-Generated Configurations}

Research on IaC quality has examined smells,
security policies, and defects~\cite{rahman2019mapping}.
Sharma et al.~\cite{sharma2016smells} investigated configuration
smells in Puppet, while Rahman et al.~\cite{rahman2019sins}
identified security smells such as hard-coded secrets and
unrestricted network access.
Saavedra and Ferreira~\cite{saavedra2022glitch} developed GLITCH
for detecting smells across IaC languages, and War et
al.~\cite{war2025sins} extended security-smell classifications
across seven IaC technologies.
Verdet et al.~\cite{verdet2025policies} examined
security policies in Terraform.
Other studies characterized IaC defects and developed
defect-prediction approaches~\cite{rahman2018bugs,
rahman2020gang,begoug2024jit,ksontini2026docker}.

Beyond security and correctness, IaC configurations can affect
cloud-resource efficiency.
Kosbar and Hamdaqa~\cite{kosbar2025smellssus} introduced
sustainability smells in Terraform, including over-provisioning,
missing autoscaling, and monolithic infrastructure.
Recent work has also evaluated LLM-generated IaC configurations:
Kon et al.~\cite{kon2024iaceval} introduced IaC-Eval;
Nekrasov et al.~\cite{nekrasov2025iacgen} identified hallucinated
Terraform resources and arguments; and Vargas et
al.~\cite{vargas2026secureterraform} examined the security
compliance of generated AWS configurations.

These studies establish the importance of IaC quality but
largely examine existing configurations or generated code in
benchmark settings. Our study instead analyzes changes proposed
in real repositories, develops an IaC-specific classification
of infrastructure operations, and compares sustainability-smell
introduction in matched agent and human PRs using the taxonomy
of Kosbar and Hamdaqa~\cite{kosbar2025smellssus}.

\section{Study Design}
\label{sec:design}
\begin{figure*}[t]
\centering
\includegraphics[width=\textwidth]{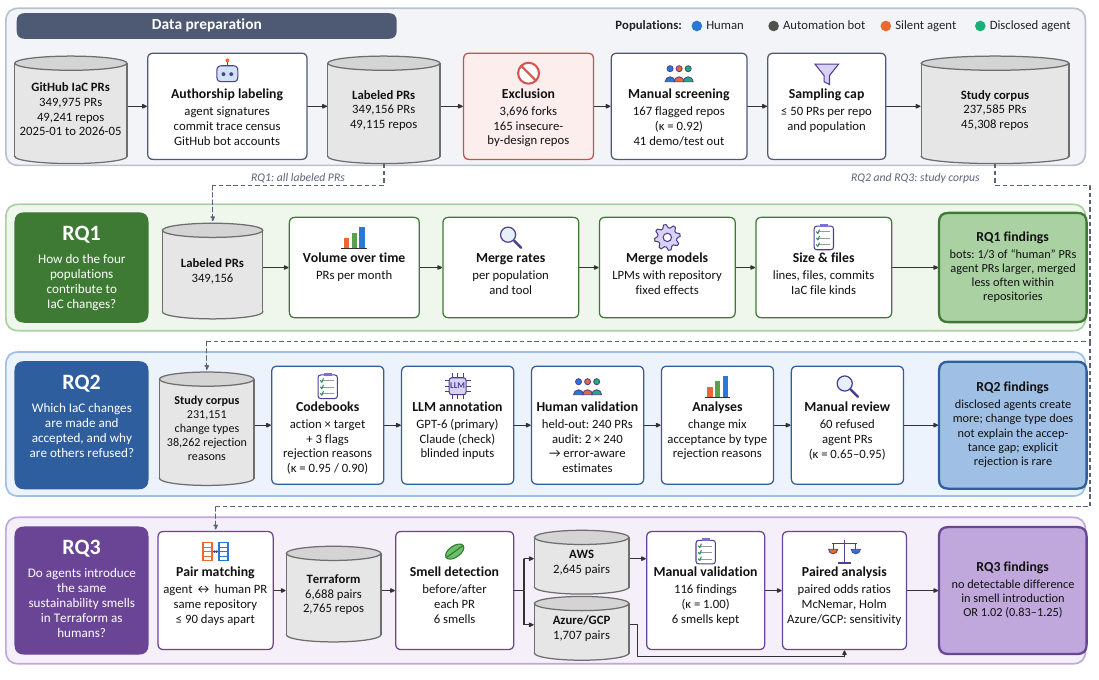}
\caption{Overview of the study design. The held-out validation covers 240 PRs (120 refused PRs coded for rejection reasons), and the audit samples 240 annotations per task. Smells are detected in all matched Terraform pairs; the main analysis uses the AWS pairs.}
\label{fig:workflow}
\end{figure*}

This section describes the data and methods shared by our
three research questions. Figure~\ref{fig:workflow} provides
an overview of the study design, while
Sections~\ref{sec:rq1}--\ref{sec:rq3} present the motivation,
approach, and results of each RQ.

\subsection{Data Collection and Authorship}
\label{sec:collection}
\textbf{PR collection.} Using the GitHub search API, we collected PRs
opened between January 1, 2025 and May 31, 2026 whose title or
description mentions a provisioning tool (Terraform, CloudFormation)
or a configuration-management tool (Ansible, Puppet, Chef); queries
were split into time windows below the API's 1,000-result limit, run
twice, and deduplicated. We keep a PR only if it changes at least one
file of these tools, identified by path and extension or, where
ambiguous, by content (e.g., \texttt{AWSTemplateFormatVersion}). This
yields 349,975 PRs from 49,241 repositories: Terraform (291,804),
Ansible (51,444), CloudFormation (7,307), Puppet (2,200), and Chef
(1,378); a PR may use several tools. For each PR, we collected its
metadata, merge status, size, commits, changed files, and review
timeline.

\textbf{Authorship labeling.} We assign each PR to one of four
populations in three steps. First, following AIDev~\cite{aidev}, a PR
is \emph{disclosed-agent} if an AI-agent account opened it (e.g., the
Copilot coding agent or Devin) or its metadata carries an agent
signature. Second, for the remaining 339,196 PRs, we match 14
signature rules adapted from Khosravani and Mockus~\cite{census}
against every commit: co-author trailers, generation footers (e.g.,
``Generated with Claude Code''), and agent e-mail addresses or
accounts, ignoring generic automation and mere mentions of an agent.
We consider a PR opened by a developer or a bot \emph{silent-agent}
if at least one of its commits matches these rules. Third, among PRs without agent
traces, those opened by an account that the GitHub API types as
\emph{Bot} are \emph{automation bot}, and all others are \emph{human}
(no agent trace found); 1,138 bot-opened PRs with agent traces remain
silent-agent. Excluding 819 PRs whose commits are no longer reachable
leaves 349,156 labeled PRs.

\textbf{Repository screening and sampling.} RQ1 uses 349,156
labeled PRs. For RQ2 and RQ3, we build a screened corpus that focuses
on development work and limits the influence of active
repositories. We exclude forks, repositories that are insecure by
design (e.g., copies of TerraGoat), and 41 repositories whose PRs only
test or demonstrate a tool, identified by two authors who screened
167 repositories with demo or test signals ($\kappa = 0.92$). Because
the 1\% most active repositories hold 19--36\% of a population's PRs,
we keep at most 50 PRs per repository and population, which
yields 237,585 PRs.

\subsection{Annotation of Changes and Rejection Reasons}
\label{sec:annotation}
RQ2 annotates the change type of 231,151 corpus PRs whose diff could
be retrieved (97.3\%; unreachable diffs are more often agent PRs) and
the rejection reasons of the 38,262 PRs among them that were closed
without merging.

\textbf{Codebook.} The change-type codebook has two axes
(Table~\ref{tab:codebook}): the \emph{action} (what the diff does to the
infrastructure configuration) and the \emph{target} (the resource type it
acts on, mapped from Terraform, CloudFormation, Ansible, Puppet, and
Chef resource types). Three flags cut across actions: \emph{security}
(the purpose is to reduce a security risk), \emph{destructive}
(the change destroys or forces replacement of a stateful
resource; \emph{possible} when it could without evidence that this is
intended), and \emph{permissions broadened} (judged by effective
access granted). PRs without a substantive IaC change are \emph{not
codable}. Twelve ordered decision rules resolve ties (e.g., the diff,
not the title, decides; a fix requires evidence of breakage).
Rejection reasons have seven categories: rejected by a maintainer,
discarded by the author or requester, superseded, duplicate, abandoned,
not intended to merge, and unclear. Two authors developed the codebooks
on a pilot of 100 PRs and a tuning set of 80 PRs, stratified by
population and outcome; coding the tuning set independently, they agreed with $\kappa =
0.95$ for actions and 0.90 for targets and rejection reasons before
adjudication.

\begin{table}[t]
\caption{Taxonomy of IaC changes. Action and target are independent
dimensions assigned to each codable PR.}
\label{tab:codebook}
\centering
\footnotesize
\setlength{\tabcolsep}{3pt}

\begin{adjustbox}{max width=\columnwidth}
\begin{tabular}{@{}
>{\raggedright\arraybackslash}p{0.32\columnwidth}
>{\raggedright\arraybackslash}p{0.65\columnwidth}
@{}}
\toprule
\textbf{Category} & \textbf{Definition (example)} \\
\midrule

\rowcolor{grouprow}
\multicolumn{2}{@{}l}{\textit{Action: what the change does to the infrastructure configuration}} \\
\midrule

Create & New module, stack, environment, or Ansible role \\
\specialrule{0.15pt}{0pt}{0pt}
Extend & New resources or tasks in an existing module \\
\specialrule{0.15pt}{0pt}{0pt}
Modify & Changed settings of existing resources, nothing broken (e.g., instance size) \\
\specialrule{0.15pt}{0pt}{0pt}
Fix & Repairs a failing plan or apply, error, or drift \\
\specialrule{0.15pt}{0pt}{0pt}
Upgrade & New provider, module, image, or tool version \\
\specialrule{0.15pt}{0pt}{0pt}
Migrate & Move to another platform, tool, or backend (e.g., local to remote state) \\
\specialrule{0.15pt}{0pt}{0pt}
Refactor & Restructuring without changing what is deployed (e.g., \texttt{moved} blocks) \\
\specialrule{0.15pt}{0pt}{0pt}
Remove, Revert & Deletes resources, or undoes an earlier change \\
\specialrule{0.15pt}{0pt}{0pt}
Document, Test, Format & Docs, tests, or style only \\

\midrule
\rowcolor{grouprow}
\multicolumn{2}{@{}l}{\textit{Target: type of resource acted on}} \\
\midrule

Identity \& access & IAM roles, policies, service accounts \\
\specialrule{0.15pt}{0pt}{0pt}
Network & VPCs, subnets, firewalls, load balancers, DNS \\
\specialrule{0.15pt}{0pt}{0pt}
Compute \& containers & Instances, autoscaling, Kubernetes, functions \\
\specialrule{0.15pt}{0pt}{0pt}
Storage \& data & Buckets, databases, caches, backups \\
\specialrule{0.15pt}{0pt}{0pt}
Secrets \& encryption & KMS keys, secrets, certificates \\
\specialrule{0.15pt}{0pt}{0pt}
Messaging \& integration & Queues, topics, event buses, API gateways \\
\specialrule{0.15pt}{0pt}{0pt}
Observability & Logs, monitoring, alerting \\
\specialrule{0.15pt}{0pt}{0pt}
Governance \& cost & Accounts, organization policies, tags, budgets \\
\specialrule{0.15pt}{0pt}{0pt}
Host configuration & Files, OS settings (Ansible, Puppet, Chef) \\
\specialrule{0.15pt}{0pt}{0pt}
IaC plumbing & Providers, state backends, lock files \\
\specialrule{0.15pt}{0pt}{0pt}
CI/CD, whole stack, other, none & IaC pipelines; several components; unmapped type; docs only \\

\bottomrule
\end{tabular}
\end{adjustbox}
\end{table}

\textbf{LLM annotation.} Two models from different vendors label every
PR independently: GPT-6 Luna (version 2026-09-22, Azure OpenAI) as the
primary annotator and Claude Haiku~5.5 (OpenRouter) as a check, which
reduces correlated errors and the risk that a model favors its own
vendor's agents. Each receives the codebook, the PR title and
description, the IaC diff, and the names of other changed files (for
rejection reasons, the description and timeline), with agent and bot
attribution removed and timeline participants replaced by roles, and
returns labels with supporting evidence as structured JSON. All results
use GPT-6 Luna's labels. The models agree on 82\% of actions, 74\% of
targets, and 93\% of rejection reasons; since disagreement signals
likely errors, a human audit samples agreeing and disagreeing PRs
equally and corrects the reported label shares for the errors found.

\textbf{LLM label validation.} We froze the configuration (codebook,
prompt, examples, models, parameters) and evaluated it on 240 held-out
PRs (30 per population and outcome), coded independently by two authors
(inter-coder $\kappa = 0.92$ for actions, 0.85 for targets, 0.95 for
rejection reasons) and adjudicated. An earlier held-out evaluation
prompted one rule revision; because rejection-reason rules were
unchanged, both samples inform their accuracy estimate. The primary
annotator reached 85\% accuracy for actions ($\kappa = 0.82$),
76\% for targets ($\kappa = 0.72$), and 92\% for rejection reasons.
After the evaluation, we required precision and recall of at least
0.70 with at least ten instances for drawing conclusions from a class.
This holds for creations, extensions, modifications, and upgrades;
the compute, data, identity, network, plumbing, and other targets;
and the permissions flag. Fixes, tests, the host target, and the
destructive flag (three positive cases) are reported as tentative.
Because it did not meet the criterion, the security flag is reported
only through estimates adjusted for annotation errors.

\textbf{LLM label audit.} A probability
sample of the production labels (the \emph{audit}: 480 PRs, 30 per
stratum of task, population, and agreement between the two models) was
coded by an author blind to both models' labels. Weighted to the
corpus, the primary annotator reached 88\% accuracy for actions, 81\%
for targets, and 88\% for rejection reasons. Both models, however,
labeled most PRs not meant to be merged and most maintainer rejections
as \emph{unclear}, even when they agreed, so we never compare raw label
shares. Instead, for each population and class, a stratified difference
estimator corrects the share of the primary's labels by the
design-weighted differences between human and primary labels in each
audit stratum, which provides design-unbiased estimates. In a
simulation, its nominal 95\% design-based intervals covered the truth
in only 81\% (change types) and 90\% (rejection reasons) of draws,
mainly because of rare classes; we therefore also compute conservative
intervals with the Agresti--Min adjustment~\cite{agresti2005matched}
and treat a difference between populations as established only when it
excludes zero under both. We draw no population-level conclusions for
rare classes such as removals and formatting. The coverage simulation
is in the replication package.

\subsection{Detection of Sustainability Smells}
\label{sec:smells}
We operationalize six sustainability smells adapted from Kosbar and
Hamdaqa~\cite{kosbar2025smellssus}, who derive them from cloud
providers' guidance and validate them in a practitioner survey
(Table~\ref{tab:smells}). Our detector applies each definition to the
parsed Terraform configuration, with thresholds derived from their
examples, and is tested on 42 cases.

\begin{table}[h]
\caption{Sustainability smells adapted from Kosbar and
Hamdaqa~\cite{kosbar2025smellssus} and operationalized
using static heuristics in our study.}
\label{tab:smells}
\centering
\footnotesize
\setlength{\tabcolsep}{3pt}
\renewcommand{\arraystretch}{1.12}
\begin{tabular}{@{}
>{\raggedright\arraybackslash}p{0.08\columnwidth}
>{\raggedright\arraybackslash}p{0.31\columnwidth}
>{\raggedright\arraybackslash}p{0.55\columnwidth}
@{}}
\toprule
\textbf{ID} & \textbf{Sustainability smell} &
\textbf{Detection summary} \\
\midrule
SS1 & Over-provisioning &
Large configured compute, database, or cache instances. \\
SS2 & Lack of auto-scaling &
Multiple instances with fixed capacity and no
autoscaling. \\
SS3 & Ignoring resource lifecycles &
Sized disks or volumes without a
\texttt{lifecycle} block. \\
SS4 & Excessive logging &
Long or unlimited log retention. \\
SS5 & State management &
Root configurations without a remote
state backend. \\
SS6 & Monolithic infrastructure &
Many resources defined in a single
configuration file. \\
\bottomrule
\end{tabular}
\end{table}

RQ3 pairs each agent Terraform PR with the closest-in-time human PR of
the same repository (at most 90 days apart, preferring the same size
class; each human PR used once; at most 20 pairs per repository),
yielding 6,688 pairs in 2,765 repositories (74\% in the same size
class; median gap 7 days). We detect smells and cloud providers in all
pairs; 321 pairs could not be analyzed. The main analysis uses the
2,645 pairs (1,183 repositories) in which both PRs change AWS
configurations, the provider Kosbar and Hamdaqa's rules target; the
1,707 Azure or Google Cloud pairs (63 of which also count as AWS pairs)
form a sensitivity analysis, and the remaining 2,078 pairs are
excluded.

For each PR, we detect smells in every configuration directory it
touches, before (at the merge base with the target branch) and after
(at the head commit); a smell is \emph{introduced} if it appears and
\emph{removed} if it disappears. Unreadable files are dropped from both
snapshots. We do not assess state management in repositories that use
Terragrunt, a tool that configures remote state itself, so a missing
backend in their Terraform files is not a smell. The main results
exclude example, test, demo, and LocalStack configurations and, for
state management, configurations that bootstrap the state store.

\textbf{Validation.} Two authors independently labeled up to 30
random introduced findings per smell as true or false positives,
blind to the population; all six smells were retained, with
precision of 0.86 to 1.00. They noted when the code or repository states a
justification for a confirmed smell.

\subsection{Statistical Analyses}
\label{sec:rqs}
\textbf{Contribution and acceptance (RQ1, RQ2).} We count PRs per month
and population and compare merge rates overall, per IaC tool, and per
agent tool. Linear probability models (LPMs) of the merge outcome
include population indicators (humans as reference), the logarithm of
lines and files changed, and creation-month and IaC-tool fixed effects,
first without and then with repository fixed effects, which compare
populations \emph{within} a repository; standard errors are clustered
by repository. For RQ2, change-type fixed effects and population-by-type
indicators give each population's merge difference from human PRs of
the same type; these models use codable PRs with a final outcome and
are repeated on the PRs on which both LLMs agree. We compare lines and
files per PR and lines per commit with Cliff's
$\delta$~\cite{cliff1993delta,romano2006cliff}, and the share of PRs
changing each of fifteen IaC file kinds.

\textbf{Change mix and rejection reasons (RQ2).} We compare error-aware
shares of actions, targets, and flags across populations, and
rejection reasons among refused PRs (Section~\ref{sec:annotation}).
Cram\'er's $V$ on raw labels measures the association between change
type and each population compared with humans.

\textbf{Delegation evidence (RQ2).} For each action with at least 100
PRs of an agent mode, agent acceptance is \emph{comparable} when the
95\% confidence interval of the within-repository difference excludes
a deficit larger than 5 points, and \emph{lower} when the interval
excludes zero and the estimated deficit is at least 5 points. Using
silent-agent acceptance, actions are grouped as \emph{delegate}, or as
\emph{delegate with targeted review} when at least 15\% of agent PRs
broaden permissions or may destroy stateful resources (pooling both
agent modes), or at least 5\% introduce a sustainability smell (RQ3);
actions with lower acceptance need \emph{further evidence}. The groups
do not apply to disclosed agents, whose PRs are merged less often for
nearly every action. These criteria are exploratory, were chosen after
examining the results, and are not validated thresholds.

\textbf{Review of refused agent PRs (RQ2).} For actions with lower
acceptance, two authors review 20 random refused agent PRs per action
(half silent, half disclosed; at most one per repository), coding
whether each was meant to be merged, who closed it and why, whether the
reason concerns agent-generated work, and whether the annotated action
is correct; disagreements are resolved by discussion.

\textbf{Paired smell comparison (RQ3).} Within each pair, we test
whether the agent and the human PR differ in introducing at least one
smell (McNemar's test~\cite{mcnemar1947}) and in how many they
introduce (Wilcoxon signed-rank test), overall and per smell, applying
Holm's correction~\cite{holm1979} across the six smells within each
comparison, for all, silent, and disclosed agents and for pairs in
which both PRs were merged. The effect size is the paired odds ratio
(pairs in which only the agent PR introduces a smell over pairs in
which only the human PR does) with a Wilson 95\%
interval~\cite{wilson1927}. We analyze removed smells likewise, report
smells per touched configuration because agent PRs are larger, and
account for pairs from the same repository with a repository-clustered
bootstrap (10,000 replicates). Sensitivity analyses use Kosbar and
Hamdaqa's detection rules, include example and test configurations, or
use the Azure and Google Cloud pairs.

\section{RQ1 --- IaC Contributors and Merge Outcomes}
\label{sec:rq1}

\begin{table}[t]
\caption{PR characteristics and merge outcomes by population.
$\Delta$: difference in merge probability from human PRs (percentage
points; repository-clustered standard errors). Adjusted: size, IaC
tool, month; within-repo: plus repository fixed effects.}
\label{tab:rq1_combined}
\centering
\footnotesize
\setlength{\tabcolsep}{3pt}
\begin{adjustbox}{max width=\columnwidth}
\begin{tabular}{@{}lrrrr|rrr@{}}
\toprule
& \multicolumn{4}{c|}{PR characteristics}
& \multicolumn{3}{c}{Merge difference $\Delta$} \\
\cmidrule(lr){2-5}
\cmidrule(l){6-8}
Population & PRs & Repos & \shortstack[r]{Median\\lines/PR}
& \multicolumn{1}{r|}{\shortstack[r]{Merge\\rate}}
& Unadjusted & Adjusted & \shortstack[r]{Within-\\repo} \\
\midrule
Human & 199,343 & 38,130 & 71 & 78.9\%
& -- & -- & -- \\
Automation bot & 111,192 & 6,297 & 26 & 51.8\%
& $-27.1$ (3.2) & $-26.2$ (3.6) & $-29.9$ (6.2) \\
Silent agent & 27,842 & 8,260 & 302 & 87.4\%
& $+8.5$ (2.3) & $+6.2$ (1.3) & $-4.0$ (1.2) \\
Disclosed agent & 10,779 & 3,980 & 247 & 76.3\%
& $-2.6$ (2.5) & $-4.2$ (1.8) & $-10.4$ (2.4) \\
\bottomrule
\end{tabular}
\end{adjustbox}
\end{table}

\rqpart{Motivation} AI coding agents increasingly contribute to IaC, but
their participation is difficult to characterize because
agent-generated changes may be submitted under developer identities
and conventional bots may be mistaken for human contributors. Beyond
identifying who produces IaC changes, we need to understand their
scale, the files they affect, and how maintainers respond to them.
RQ1 compares the participation, PR sizes, changed files, and merge
outcomes of humans, automation bots, silent agents, and disclosed
agents.

\rqpart{Approach} We analyze all 349,156 labeled PRs
(Section~\ref{sec:collection}). For participation, we count PRs per
month and population; for size, we compare lines and files changed
per PR and lines per commit; for changed files, we measure the share
of PRs that touch each of fifteen IaC file kinds; and for merge
outcomes, we compare merge rates with those of human PRs, first
unadjusted, then adjusted for PR size, IaC tool, and month, and
finally within repositories (Section~\ref{sec:rqs}).

\rqresults

\begin{figure}[t]
\centering
\includegraphics[width=0.85\columnwidth]{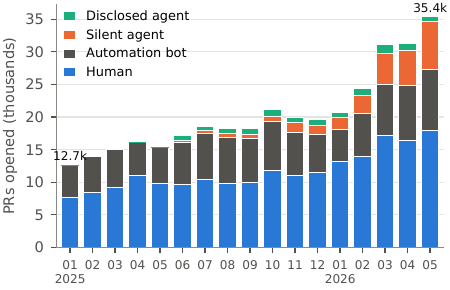}
\caption{IaC PRs opened per month, by population.}
\label{fig:rq1monthly}
\vspace{-0.5cm}
\end{figure}

\textbf{Bots and silent agents make up a large and growing share
of IaC PRs.}
Of the 349,156 PRs, 57.1\% are authored by humans, 31.8\% by
automation bots, 8.0\% by silent agents, and 3.1\% by disclosed
agents: bots open about a third of the PRs that PR metadata
labels as human, and silent-agent PRs outnumber disclosed-agent
PRs 2.6 to 1. Monthly IaC PR volume almost tripled over the study
period, from 12.7k PRs in January 2025 to 35.4k in May 2026
(Figure~\ref{fig:rq1monthly}). Human and bot PRs grew moderately,
whereas agent PRs rose from fewer than 100 per month before May
2025 to 8,057 in May 2026; disclosed-agent PRs appeared in June
2025 and stayed between about 700 and 1,300 per month, while
silent-agent PRs kept growing and outnumber them since November
2025.

\begin{figure}[b]
\centering
\includegraphics[width=\columnwidth]{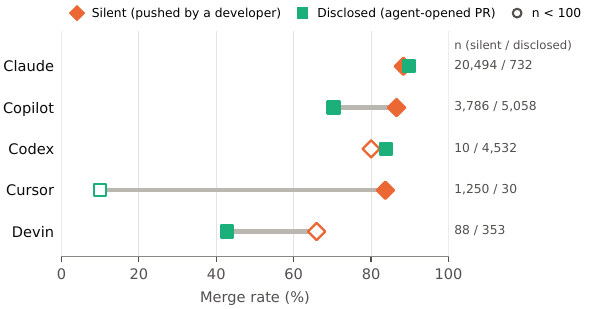}
\caption{Agent PR merge rates by tool and attribution mode. Tools seen
in only one mode and 1,967 silent PRs of unknown tool are omitted.}
\label{fig:rq1acceptance}
\end{figure}

\textbf{Silent agents lead in overall merge rates, but both agent
modes lag humans within repositories.}
Silent-agent PRs have the highest overall merge rate (87.4\%),
compared with 78.9\% for human PRs, 76.3\% for disclosed-agent
PRs, and 51.8\% for bot PRs, consistently for Terraform, Ansible,
and CloudFormation. However, raw merge rates do not account for
differences across projects. Accounting for PR size, IaC tool,
and creation month reduces the silent-agent advantage over humans
from 8.5 to 6.2 percentage points. Further accounting for
repository differences reverses this association
(Table~\ref{tab:rq1_combined}): silent-agent PRs are merged
4.0 points and disclosed-agent PRs 10.4 points less often than
human PRs within the same repositories. The overall lead of
silent agents is consistent with where they are used: even human
PRs are merged at 86.2\% in repositories with silent-agent PRs,
compared with 75.6\% elsewhere. Bot PRs are merged 26 to 30
points less often under every specification, mostly because
superseded dependency bumps are closed.

\textbf{Agent-tool merge rates vary by attribution mode}
(Figure~\ref{fig:rq1acceptance}). Copilot PRs are merged at
86.6\% when a developer pushes the agent's commits (3,786 PRs)
and at 70.3\% when the agent opens the PR (5,058); Devin shows
the same pattern (65.9\% vs.\ 42.8\%, with only 88 silent PRs),
while Claude is merged at similar rates in both forms (88.3\% and
89.8\%).

\textbf{Agent PRs are larger and focus on infrastructure modules,
while bots maintain dependencies.}
The median silent-agent PR changes 302 lines in 8 files and
the median disclosed-agent PR 247 lines in 7 files, compared
with 71 lines in 4 files for humans and 26 lines in one file
for bots (Table~\ref{tab:rq1_combined}); the effect sizes are
small for agents (Cliff's $\delta = 0.31$ and $0.28$ on lines)
and medium for bots ($-0.38$). Agents deliver more code per
PR rather than more commits: all three non-bot populations
typically use two commits per PR.

\begin{figure}[h]
\centering
\includegraphics[width=\columnwidth]{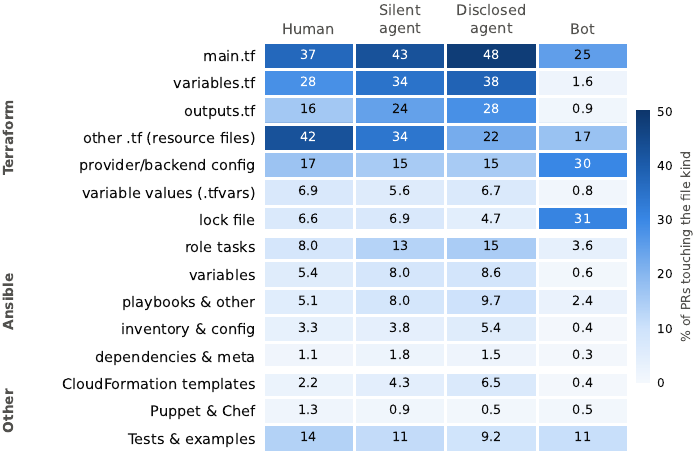}

\caption{Share of population PRs that change at least
one file of each kind.}

\label{fig:rq1files}
\end{figure}

The file-level comparison (Figure~\ref{fig:rq1files}) shows
that agents modify Terraform module skeletons more often
than humans: \texttt{main.tf} (43\% of silent-agent and 48\%
of disclosed-agent PRs vs.\ 37\% of human PRs),
\texttt{variables.tf} (34\% and 38\% vs.\ 28\%), and
\texttt{outputs.tf} (24\% and 28\% vs.\ 16\%). Humans more
often change other \texttt{.tf} resource files (42\% vs.\
34\% and 22\%). In Ansible, agents change role tasks more
often than humans (13--15\% vs.\ 8\%). Bots concentrate on
dependencies: 31\% of bot PRs change the Terraform lock
file and 30\% change provider or backend configurations.

\begin{tcolorbox}[
colback=gray!8,
colframe=gray!60,
boxrule=0.5pt,
left=4pt,
right=4pt,
top=3pt,
bottom=3pt]

\textbf{Summary of RQ1.} 
Bots open about a third of ostensibly human PRs, while silent-agent
contributions grow rapidly. Both agent modes lag humans in
within-repository merge rates despite silent agents leading overall.
Copilot and Devin PRs are merged more often when developer-pushed.
Agent PRs are about four times larger and often change Terraform
module skeletons; bots maintain dependencies.

\end{tcolorbox}

\section{RQ2 --- Changes and Acceptance}
\label{sec:rq2}

\begin{table}[t]
\caption{Error-aware share of labeled PRs per population (\%). Bold:
difference from humans excludes zero; $^{\circ}$: not under the
conservative intervals; $^{\dagger}$: tentative class.}
\label{tab:rq2mix}
\centering
\footnotesize
\renewcommand{\arraystretch}{0.94}
\begin{tabular*}{\columnwidth}{@{}c@{\extracolsep{\fill}}lrrrr@{}}
\toprule
 & & Human & Bot & Silent & Disclosed \\
\midrule
\multirow{11}{*}{\rotatebox[origin=c]{90}{\emph{Action}}} & Create & 21.4 & \textbf{0.5} & 19.9 & \textbf{34.0} \\
 & Extend & 20.2 & \textbf{0.8} & 18.8 & \textbf{7.5} \\
 & Modify & 19.3 & \textbf{2.0} & 14.7 & 11.2 \\
 & Fix$^{\dagger}$ & 8.6 & \textbf{0.0} & 15.4 & 16.1 \\
 & Upgrade & 7.9 & \textbf{88.2} & \textbf{2.7}$^{\circ}$ & \textbf{2.6}$^{\circ}$ \\
 & Migrate$^{\dagger}$ & 1.1 & 0.3 & 4.4 & 4.5 \\
 & Refactor$^{\dagger}$ & 3.7 & \textbf{0.1}$^{\circ}$ & 5.2 & \textbf{4.2}$^{\circ}$ \\
 & Remove$^{\dagger}$ & 4.7 & \textbf{0.2}$^{\circ}$ & 3.7 & 1.8 \\
 & Document$^{\dagger}$ & 6.5 & \textbf{4.3}$^{\circ}$ & 6.5 & \textbf{3.3}$^{\circ}$ \\
 & Test$^{\dagger}$ & 6.7 & \textbf{2.3}$^{\circ}$ & 4.6 & 8.2 \\
 & Format$^{\dagger}$ & 0.2 & 0.6 & 0.7 & 3.5 \\
\midrule
\multirow{7}{*}{\rotatebox[origin=c]{90}{\emph{Target}}} & Compute & 26.0 & \textbf{7.1} & 25.2 & 23.3 \\
 & IaC plumbing & 13.5 & \textbf{78.5} & 9.0 & 13.4 \\
 & Network & 11.8 & \textbf{4.2}$^{\circ}$ & 9.1 & 14.3 \\
 & Storage \& data & 8.5 & \textbf{2.2}$^{\circ}$ & 9.6 & 14.1 \\
 & Identity \& access & 7.6 & \textbf{2.4}$^{\circ}$ & \textbf{10.8}$^{\circ}$ & \textbf{5.8}$^{\circ}$ \\
 & Host config.$^{\dagger}$ & 6.3 & \textbf{0.4}$^{\circ}$ & 8.9 & 5.9 \\
 & Observability$^{\dagger}$ & 8.1 & \textbf{1.2}$^{\circ}$ & 8.0 & 5.4 \\
\midrule
\multirow{3}{*}{\rotatebox[origin=c]{90}{\emph{Flag}}} & Permissions broad. & 9.8 & \textbf{0.2} & 14.2 & 9.5 \\
 & Destructive$^{\dagger}$ & 8.5 & \textbf{0.3}$^{\circ}$ & 4.5 & 1.6 \\
 & Security & 5.4 & \textbf{0.5}$^{\circ}$ & 8.8 & \textbf{10.3}$^{\circ}$ \\
\bottomrule
\end{tabular*}
\vspace{-0.3cm}

\end{table}

\rqpart{Motivation} RQ1 establishes who contributes to IaC and
how often their PRs are merged, but merge rates do not
reveal what changes contributors propose or why some PRs
are not merged. Prior work has examined failed and rejected
agent PRs~\cite{nakashima2026rejected},
but how acceptance varies across infrastructure-specific
changes remains unclear. Creating resources, extending modules,
changing permissions, and upgrading dependencies involve different
operations and risks. These differences raise questions
about which IaC changes humans, automation bots, and agents
propose, how acceptance varies by change type, and why
some contributions are closed without merging.

\rqpart{Approach} Two LLMs from different vendors annotate
231,151 PRs with an action, target, and risk flags, and
38,262 refused PRs with rejection reasons. A stratified
human audit corrects label shares for annotation error
(Section~\ref{sec:annotation}). We compare error-aware
change and rejection-reason shares across populations and
within-repository merge rates adjusted for action. We manually
review 60 refused agent PRs from lower-acceptance
actions and assess task delegation using exploratory
acceptance and risk criteria (Section~\ref{sec:rqs}).
We consider population differences reliable when
design-based and conservative confidence intervals
indicate a difference.

\rqresults

\begin{table*}[t]

\caption{Exploratory IaC delegation evidence. Human: merge rate;
$\Delta$: within-repository merge difference from humans (points,
95\% CI; ${\geq}137$ PRs per population). Perm./Destr.: pooled agent
PRs broadening permissions or potentially destroying stateful
resources. Smells: agent/human PRs in matched RQ3 pairs introducing
smells (--: $n<40$). Groups follow Section~\ref{sec:rqs}, using silent-agent
acceptance and pooled risk indicators; italics: tentative
($^{\dagger}$, Destr.); $^{b}$: acceptance inconclusive.}
\label{tab:rq2guide}
\centering
\footnotesize
\setlength{\tabcolsep}{4pt}
\begin{tabular*}{\textwidth}{@{}l@{\extracolsep{\fill}}rrrrrrl@{}}
\toprule
Action & Human (\%) & Silent $\Delta$ & Disclosed $\Delta$ &
Perm. (\%) & Destr. (\%) & Smells (\%) & Group \\
\midrule
Modify & 90.1 & $-$0.6 [$-$2.0, +0.9] & $-$9.3 [$-$13.1, $-$5.4] & 10.4 & 7.5 & 0.9 / 2.7 & Delegate \\
Fix$^{\dagger}$ & 90.3 & $-$1.2 [$-$2.7, +0.3] & $-$6.1 [$-$9.9, $-$2.2] & 11.9 & 4.2 & 0.5 / 1.0 & \textit{Delegate} \\
Format$^{\dagger}$ & 85.4 & +0.0 [$-$4.7, +4.8] & $-$6.9 [$-$12.8, $-$1.0] & 0.0 & 0.0 & -- & \textit{Delegate} \\
\midrule
Create & 86.8 & $-$0.7 [$-$2.6, +1.2] & $-$10.1 [$-$13.6, $-$6.5] & 10.5 & 2.5 & 39.4 / 39.8 & Targeted review \\
Extend & 88.7 & $-$0.4 [$-$2.1, +1.3] & $-$10.4 [$-$13.9, $-$6.9] & 24.1 & 5.6 & 6.7 / 9.8 & Targeted review \\
Migrate$^{\dagger}$ & 89.7 & +0.8 [$-$1.2, +2.8] & $-$9.1 [$-$14.0, $-$4.1] & 12.3 & 25.3 & 12.0 / 14.9 & \textit{Targeted review} \\
Refactor$^{\dagger}$ & 91.2 & +0.1 [$-$2.3, +2.4] & $-$11.2 [$-$18.9, $-$3.4] & 2.0 & 15.2 & 14.9 / 8.9 & \textit{Targeted review} \\
Remove$^{\dagger}$ & 91.8 & $-$2.3 [$-$5.2, +0.5] & $-$7.8 [$-$13.4, $-$2.2] & 1.7 & 27.9 & 3.3 / 1.8 & \textit{Targeted review$^{b}$} \\
\midrule
Upgrade & 89.6 & $-$11.6 [$-$15.1, $-$8.1] & $-$12.2 [$-$19.2, $-$5.3] & 1.1 & 3.4 & 0.0 / 1.4 & Further evidence needed \\
Document$^{\dagger}$ & 89.3 & $-$6.5 [$-$10.1, $-$2.9] & $-$10.7 [$-$17.6, $-$3.8] & 0.4 & 0.1 & 0.0 / 0.0 & \textit{Further evidence needed} \\
Test$^{\dagger}$ & 85.1 & $-$9.6 [$-$18.4, $-$0.7] & +2.7 [$-$22.8, +28.2] & 0.3 & 0.6 & -- & \textit{Further evidence needed} \\
\bottomrule
\end{tabular*}
\vspace{-0.3cm}
\end{table*}

\textbf{No robust change-type differences between silent
agents and humans were established; disclosed agents
concentrate on creating new infrastructure.}
Silent-agent PRs resemble human PRs: the association between
population and action is weak (Cram\'er's $V = 0.12$;
0.07 for targets), and no difference from humans is
established (Table~\ref{tab:rq2mix}). Disclosed agents create
new modules, stacks, or environments more often than humans
(34.0\% vs.\ 21.4\%; +12.6 points, 95\% CI 4.7--20.4)
and extend existing infrastructure less often
(7.5\% vs.\ 20.2\%; $-$12.7, $-$22.0 to $-$3.4).
Bots differ substantially: 88.2\% of their PRs are upgrades
and 78.5\% act on IaC plumbing. Point estimates suggest
that agent PRs are fixes about twice as often as human PRs
and silent-agent PRs broaden permissions more often
(14.2\% vs.\ 9.8\%), but neither difference is established.

\begin{figure}[t]
\centering
\includegraphics[width=0.9\columnwidth]{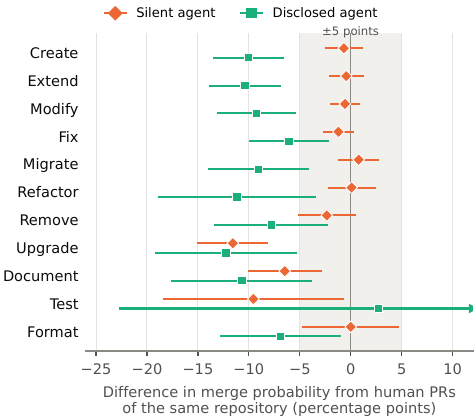}
\caption{Within-repository difference in merge probability between
agent and human PRs by action (repository-fixed-effects LPM; repository-clustered
95\% CIs; at least 100 PRs per agent mode). Shaded: $\pm$5 points.}
\label{fig:rq2accept}
\vspace{-0.3cm}
\end{figure}

\textbf{Change type does not explain the lower within-repository
merge rates of agent PRs.}
Among 211,985 codable PRs with final outcomes, silent-agent
PRs are merged 2.1 points less often than human PRs within
repositories (95\% CI $-$3.4 to $-$0.8), and disclosed-agent
PRs 8.4 points less often ($-$12.3 to $-$4.5).
Adjusting for action leaves both gaps unchanged. These
estimates differ from RQ1 because they use the screened,
capped corpus.

Figure~\ref{fig:rq2accept} and Table~\ref{tab:rq2guide}
show differences by action. Silent-agent deficits concentrate
in upgrades ($-$11.6 points, 95\% CI $-$15.1 to $-$8.1),
tests, and documentation; for creations, extensions,
modifications, fixes, migrations, and refactorings, intervals
exclude deficits larger than 3 points. Disclosed-agent PRs
are merged 6.1--12.2 points less often across actions,
except tests, whose estimate is inconclusive.
Across repositories, silent-agent PRs are merged 1.6 points
more often (95\% CI 0.7--2.4), consistent with
concentration in repositories with higher human merge rates
(RQ1). By target, silent-agent deficits remain within
3 points except IaC plumbing ($-$4.8,
95\% CI $-$7.0 to $-$2.6); estimates for identity and access
($-$1.7, $-$3.6 to 0.3) and permission broadening
($-$1.6, $-$3.5 to 0.3) are small.
Among PRs where both LLMs agree, the upgrade deficit
persists ($-$14.2 points for all agents), but intervals for
other actions include zero. Observational
comparisons do not establish causes: PR quality, developer
experience, and review practices may differ within the
same repository.

\textbf{Explicit maintainer rejection is rarely recorded among
refused IaC PRs.}
Table~\ref{tab:rq2reject} shows error-aware rejection-reason
shares by population. Explicit maintainer rejection accounts
for 1.4\% of refused bot PRs to 6.4\% of disclosed-agent PRs;
only 19 such cases appear in the audit, and conservative
intervals extend to 8--12\%, preventing reliable comparisons
across populations for this and other rare reasons.
Most refused bot PRs are superseded by newer versions (69.6\%).
Among refused human PRs, 28.7\% (95\% CI 14.5--42.9)
were not intended to merge, compared with 3.4\% of silent-agent
and 0.6\% of disclosed-agent PRs; uncorrected labels
underestimate the human share at 5.3\%.
Most refused agent PRs have no stated closure reason
(54.0\% silent; 75.4\% disclosed). Silent-agent PRs are also
superseded more often than human PRs (21.7\% vs.\ 12.8\%;
+8.9 points, 95\% CI 6.5--11.3).

\begin{table}[b]
\vspace{-0.6cm}
\caption{Error-aware share of refused PRs per reason (\%; raw share of
the primary's labels in parentheses). Bold and $^{\circ}$ as in
Table~\ref{tab:rq2mix}.}
\label{tab:rq2reject}
\centering
\footnotesize
\setlength{\tabcolsep}{2.5pt}
\begin{tabular}{@{}lrrrr@{}}
\toprule
 & Human & Bot & Silent & Disclosed \\
\midrule
Rejected by maintainer & 5.0 (4.2) & \textbf{1.4 (1.3)}$^{\circ}$ & \textbf{2.8 (3.7)}$^{\circ}$ & 6.4 (7.3) \\
Discarded by author & 7.0 (7.8) & \textbf{3.4 (3.8)}$^{\circ}$ & \textbf{10.7 (10.4)}$^{\circ}$ & \textbf{2.5 (2.5)}$^{\circ}$ \\
Superseded & 12.8 (14.1) & \textbf{69.6 (68.8)} & \textbf{21.7 (21.1)} & 10.8 (7.5) \\
Duplicate & 1.7 (1.2) & \textbf{0.0 (0.2)}$^{\circ}$ & 1.2 (1.2) & 0.9 (0.9) \\
Abandoned & 2.9 (3.3) & \textbf{1.7 (1.9)}$^{\circ}$ & \textbf{6.2 (6.2)}$^{\circ}$ & 3.4 (3.4) \\
Not meant to merge & 28.7 (5.3) & \textbf{0.0 (0.0)} & \textbf{3.4 (2.4)} & \textbf{0.6 (0.6)} \\
Unclear & 41.9 (64.0) & \textbf{23.8 (23.8)} & 54.0 (54.9) & \textbf{75.4 (77.8)} \\
\midrule
Refused PRs & 14,778 & 20,482 & 2,012 & 990 \\
\bottomrule
\end{tabular}
\end{table}

\textbf{Delegation evidence varies by change type and risk,
while many refused agent upgrades reflect cleanup or replacement.}
Table~\ref{tab:rq2guide} applies the exploratory criteria.
\emph{Delegate:} modifications and, tentatively, fixes and
formatting show comparable silent-agent acceptance and remain
below risk thresholds.
\emph{Targeted review:} creations, extensions, and, tentatively,
migrations and refactorings show comparable acceptance but
carry risk indicators. Removals are also placed here, although
their acceptance evidence is inconclusive. Creations introduce
sustainability smells in 39.4\% of agent and 39.8\% of human PRs
(mostly missing remote state); 24.1\% of agent extensions
broaden permissions, while roughly a quarter of agent
migrations and removals may destroy stateful resources.
\emph{Further evidence needed:} agent upgrades and, tentatively,
documentation and test changes are merged less often.

A manual review of 60 refused agent PRs from
lower-acceptance actions (pre-adjudication $\kappa=0.65$--$0.95$)
found 44 were intended to merge. Closure reasons were
unavailable for 27, replacement PRs accounted for 12,
objections to the approach or failing checks preceded 6,
and only 4 closures, all disclosed-agent PRs, explicitly
concerned agent-generated work. Among 20 upgrades,
18 were correctly classified; four of ten silent-agent
upgrades were closed as stale by automation and three
superseded by newer updates, suggesting cleanup rather than
rejection on the merits. In contrast, only 7 of 20 documentation
PRs and 7 of 20 test PRs were correctly classified, often
because they accompanied feature work; their acceptance
estimates cannot reliably characterize those tasks.

\begin{tcolorbox}[colback=gray!8,colframe=gray!60,boxrule=0.5pt,left=4pt,right=4pt,top=3pt,bottom=3pt]

\textbf{Summary of RQ2.} No robust change-type differences
were established between silent agents and humans; disclosed
agents create more infrastructure, while bots favor upgrades.
Within repositories, silent-agent PRs show comparable acceptance
for most actions, whereas disclosed-agent PRs are merged
6--12 points less often; change type does not explain this gap.
Explicit maintainer rejection is rare. Modifications show
comparable acceptance and few risk indicators, while creations,
extensions, migrations, and removals warrant targeted checks
for smells, permissions, and destruction. Many refused agent
upgrades reflect automated cleanup or replacement; documentation
and test findings rely on unreliable labels.

\end{tcolorbox}

\section{RQ3 --- Agent vs. Human Sustainability Smells}

\label{sec:rq3}

\rqpart{Motivation} RQ2 examines the types of IaC changes and
their acceptance, but merge decisions do not reveal whether
the resulting configurations contain resource-efficiency
concerns. Infrastructure choices such as over-provisioning,
missing autoscaling, and unmanaged state can persist beyond
a PR's merge. Kosbar and Hamdaqa~\cite{kosbar2025smellssus}
characterize such patterns as sustainability smells.
Whether agent PRs introduce these smells more often than
human PRs within the same repositories remains unclear.

\begin{table*}[t]
\caption{Sustainability smells introduced in the 2,645 matched AWS pairs.
Precision: share of sampled findings confirmed by both authors ($n$).
Agent, Human: \% of PRs introducing the smell. A/H: pairs in which only
the agent or only the human PR does. OR: paired odds ratio, 95\% CI (--:
no such pairs). No smell differs (McNemar, Holm-adjusted $p = 1.00$ per
smell; any of the six: $p = 0.92$).}
\label{tab:rq3smells}
\centering
\footnotesize
\setlength{\tabcolsep}{3pt}
\begin{tabular*}{\textwidth}{@{}l@{\extracolsep{\fill}}rrrrrrrrr@{}}
\toprule
 & & & & \multicolumn{2}{c}{All agents (2,645)} & \multicolumn{2}{c}{Silent (2,294)} & \multicolumn{2}{c}{Disclosed (351)} \\
\cmidrule(lr){5-6}\cmidrule(lr){7-8}\cmidrule(l){9-10}
Smell & Precision ($n$) & Agent (\%) & Human (\%) & A/H & OR [95\% CI] & A/H & OR [95\% CI] & A/H & OR [95\% CI] \\
\midrule
SS1 Over-provisioning & 1.00 (13) & 0.08 & 0.15 & 2/4 & 0.50 [0.11, 2.33] & 2/3 & 0.67 [0.13, 3.33] & 0/1 & 0.00 [0.00, 3.84] \\
SS2 Lack of auto-scaling & 0.86 (7) & 0.19 & 0.08 & 5/2 & 2.50 [0.56, 11.16] & 5/2 & 2.50 [0.56, 11.16] & 0/0 & -- \\
SS3 Ignoring lifecycles & 1.00 (6) & 0.11 & 0.11 & 3/3 & 1.00 [0.23, 4.33] & 2/1 & 2.00 [0.26, 15.26] & 1/2 & 0.50 [0.07, 3.82] \\
SS4 Excessive logging & 1.00 (30) & 0.91 & 0.95 & 20/21 & 0.95 [0.52, 1.74] & 19/20 & 0.95 [0.51, 1.76] & 1/1 & 1.00 [0.10, 9.58] \\
SS5 State management & 0.97 (30) & 7.71 & 7.64 & 133/131 & 1.02 [0.80, 1.29] & 101/107 & 0.94 [0.72, 1.24] & 32/24 & 1.33 [0.79, 2.25] \\
SS6 Monolithic & 1.00 (30) & 2.12 & 2.31 & 42/47 & 0.89 [0.59, 1.35] & 36/38 & 0.95 [0.60, 1.49] & 6/9 & 0.67 [0.25, 1.80] \\
\midrule
Any of the six & 0.98 (116) & 10.36 & 10.25 & 179/176 & 1.02 [0.83, 1.25] & 141/145 & 0.97 [0.77, 1.23] & 38/31 & 1.23 [0.77, 1.96] \\
\bottomrule
\end{tabular*}
\vspace{-0.3cm}
\end{table*}

\rqpart{Approach} We match each agent PR to a human PR
from the same repository and period, detect sustainability
smells in every Terraform configuration a PR touches before
and after the change, and compare smell introduction within
pairs using paired odds ratios (OR) and McNemar tests
(Section~\ref{sec:smells}). We analyze 12,737 of 13,376
matched PRs (95.2\%), mostly losing PRs whose repositories
can no longer be cloned. The main analysis covers the 2,645
pairs from 1,183 repositories in which both PRs change AWS
configurations (2,294 silent-agent and 351 disclosed-agent
pairs), including 2,039 pairs in which both PRs were merged.
Parse failures and unavailable repositories affect the
populations at similar rates. Sensitivity analyses use the
1,707 Azure and Google Cloud pairs, the published detection
rules, and a repository-clustered bootstrap.

\rqresults

\textbf{Agent PRs introduce sustainability smells as often as human
PRs from the same repository and period.} An agent PR introduces at
least one smell in 10.4\% of the pairs and its human counterpart in
10.3\% (paired OR 1.02, 95\% CI 0.83--1.25; McNemar $p = 0.92$;
Table~\ref{tab:rq3smells}). The number of introduced smells is also
similar (400 vs.\ 416; Wilcoxon $p = 0.99$), and agents introduce
slightly fewer per touched configuration (0.055 vs.\ 0.064). The result
holds for silent agents (9.6\% vs.\ 9.8\%; OR 0.97, 0.77--1.23) and for
pairs in which both PRs were merged (8.9\% vs.\ 8.8\%; OR 1.01,
0.79--1.29). For disclosed agents, the estimate is less precise
(15.1\% vs.\ 13.1\%; OR 1.23, 0.77--1.96), and for their merged pairs
(242 pairs; OR 1.56, 0.87--2.79) the interval is too wide to rule out
no difference or a substantial excess. Smells cluster in repositories:
both PRs of a pair introduce a smell in 95 pairs, 3.4 times the 28
expected if smells occurred by chance.

\textbf{No single smell differs between agents and humans; most are
missing remote state.} No smell shows a detectable difference
(smallest unadjusted $p = 0.45$; Table~\ref{tab:rq3smells}). The most
common smell is State management (SS5): a root configuration without a
remote backend for its state. It appears in 7.7\% of agent and 7.6\% of
human PRs and makes up 73\% and 67\% of the smells they introduce.
Monolithic configurations (2.1\% vs.\ 2.3\%) and excessive logging
(0.9\% vs.\ 1.0\%) follow; the other three appear in fewer than 0.2\%
of PRs.

\textbf{Agents and humans introduce more smells than they remove.}
Agent PRs introduce 400 smells and remove 166, a net increase of 234;
human PRs introduce 416 and remove 204, a net increase of 212. Removing
smells is rare for both: an agent PR removes at least one smell in
3.7\% of the pairs and its human counterpart in 4.5\%, with no
statistically detectable difference (paired OR 0.81, 95\% CI
0.60--1.08; $p = 0.18$).

\begin{figure}[b]
\vspace{-0.5cm}

\centering
\includegraphics[width=\columnwidth]{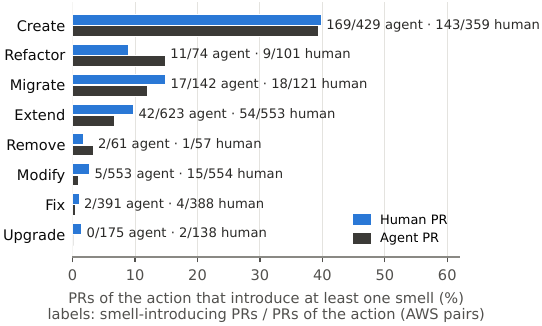}
\caption{Share of agent and human PRs (AWS pairs) introducing a smell,
by RQ2 action (${\geq}50$ PRs per side). Labels: smell-introducing / all PRs.}
\label{fig:rq3action}
\end{figure}

\textbf{Most smells arise from creating or restructuring
configurations, for agents and humans alike.} Creating new
infrastructure is the riskiest action: 39.4\% of agent and 39.8\% of
human creations introduce a smell (Figure~\ref{fig:rq3action}).
Restructuring follows, with migrations (12.0\% and 14.9\%),
refactorings (14.9\% and 8.9\%), and extensions (6.7\% and 9.8\%),
whereas modifications, fixes, and upgrades rarely introduce smells
(at most 2.7\%). Creations alone account for 62\% of the agent and 53\%
of the human PRs that introduce a smell. Comparing agent and human PRs
of the same action shows no detectable difference ($-$1.0 points,
95\% CI $-$2.5 to 0.6).

\textbf{The detector is precise, and a quarter of confirmed smells
come with a stated reason.} Two authors checked a sample of 116
detected smells and confirmed 114 (precision 0.98, 95\% CI 0.94--1.00).
Not every confirmed smell is a mistake: for 28 of the 114 (25\%), the
code or repository explains why the configuration was chosen. This
holds for agent PRs (12 of 56) and human PRs (16 of 58) alike. Examples
include 11 of the 13 over-provisioned resources, such as large
benchmark worker pools, and one-year log retention required for
compliance.

\textbf{No statistically detectable difference appears across
sensitivity analyses.} We repeated the main comparison in four ways.
Including example and test configurations gives OR 1.14 (0.95--1.38);
using the published detection rules instead of our detector gives
OR 1.10 (0.93--1.29); and analyzing the 1,707 Azure and Google Cloud
pairs gives OR 1.27 (0.95--1.69). A repository-clustered bootstrap,
which accounts for repositories contributing several pairs, widens the
interval only slightly (0.80--1.28 instead of 0.83--1.25). Of the 42
sensitivity comparisons of introducing any smell, three reach
$p < 0.05$ before correction, all on Azure and Google Cloud; none
remains significant after correcting for multiple tests.

\begin{tcolorbox}[colback=gray!8,colframe=gray!60,boxrule=0.5pt,left=4pt,right=4pt,top=3pt,bottom=3pt]
\textbf{Summary of RQ3.} Across 2,645 matched pairs, agent PRs
introduce sustainability smells no more often than human PRs from the
same repository and period (10.4\% vs.\ 10.3\%; paired OR 1.02, 95\% CI
0.83--1.25), including silent agents and merged pairs; the
disclosed-agent estimate is less precise. Smells depend on the task and
the repository rather than on who writes the change: most are missing
remote state, and they arise mainly when infrastructure is created.
Sustainability checks should therefore target creation PRs and state
configuration, whether an agent or a human authored them.
\end{tcolorbox}

\section{Discussion}
\label{sec:discussion}
Our findings show that contributor provenance, maintainer acceptance,
and sustainability smells capture distinct aspects of AI-agent
participation in IaC. PR authorship alone misses silent-agent
contributions and misclassifies automation bots, while aggregate merge
rates obscure repository differences. Within repositories, disclosed-agent
PRs are merged less often for nearly every change type, whereas
silent-agent PRs are not; visible agent authorship may attract closer
scrutiny, or disclosed agents may work with less developer oversight.
Developer-pushed agent changes are merged more often than agent-opened
PRs for some tools, potentially reflecting developer selection rather
than submission mode. Change type does not explain the acceptance gap,
and PR closure does not necessarily indicate rejection of agent work.
Studies should therefore detect silent agents and not evaluate agents
by PR metadata or merge rates alone.

For practitioners, agent-generated modifications appear promising for
delegation, while creations, extensions, migrations, and removals
warrant targeted checks of resource provisioning, permissions, and
destructive changes. Sustainability smells occur at comparable rates in
matched agent and human PRs, mostly when infrastructure is created, and
some have stated justifications rather than indicating waste. Because
missing remote state accounts for most of them, a repository-level CI
check for a remote backend would catch most smells regardless of
authorship. Tool vendors can improve transparency through
machine-readable agent provenance and clearer explanations of proposed
changes.
\section{Threats to Validity}
\label{sec:threats}
\textbf{Construct validity.} Agent use leaving no trace in PRs or
commits is missed, as are bots running under user accounts, so
silent-agent counts are lower bounds. Change types and rejection
reasons are LLM labels; we froze the configuration, validated it on
held-out PRs, and corrected population shares with a stratified human
audit. Because the audit is small for rare classes, we draw conclusions
only where differences hold under conservative intervals. Merge
outcomes measure acceptance, not quality, and smells indicate potential
waste rather than measured cost or energy. Our detector adapts
published smell definitions; its precision is 0.98 (recall is
unknown), and the published rules yield the same result.

\textbf{Internal validity.} Our comparisons are observational. To limit
confounding, we compare PRs within repositories, match agent and human
PRs by repository and period, and, because agent PRs are larger, report
smells per configuration and by change type. Repository-clustered
standard errors and a bootstrap account for non-independent PRs. We
interpret confidence intervals rather than non-significant tests as
evidence of similarity. Diffs were unavailable for 2.7\% of PRs, more
often agent PRs, which could bias RQ2 and RQ3 if these differ
systematically. Undetected agent use among human PRs would bias
comparisons towards no difference. Delegation criteria were set
after examining results and are therefore exploratory.

\textbf{External validity.} We study public GitHub repositories from
January 2025 to May 2026, dominated by Terraform, and RQ3 focuses on
AWS. Azure and Google Cloud pairs point in the same direction, but
private deployments, other tools, and later agent versions may differ.
\section{Conclusion}
\label{sec:conclusion}

AI coding agents increasingly contribute to Infrastructure as Code,
where changes affect deployed cloud resources, yet their
contributions are often difficult to identify. This study analyzed
349,156 IaC pull requests from 49,241 GitHub repositories to examine
who proposes such changes, which changes are accepted, and which
sustainability smells they introduce.

Our results show three main findings. First, PR metadata does not
reliably reflect contributor identity: automation bots open about a
third of PRs attributed to humans, and a substantial share of agent
contributions is undisclosed. Second, acceptance depends on how agent
work is submitted: within repositories, silent-agent PRs are accepted
at rates comparable to human PRs for most change types, whereas
disclosed-agent PRs are merged 6 to 12 points less often, a gap that
change type does not explain. Third, agent and human PRs introduce
sustainability smells at comparable rates (10.4\% vs.\ 10.3\%), with no
statistically detectable difference; smells arise predominantly when
new infrastructure is created and reflect project conventions rather
than authorship.

These findings suggest that empirical studies should account for
undisclosed agent contributions and that maintainers should apply
project-wide IaC checks independently of authorship. Future work should
relate IaC changes to deployed resources and their operational cost.

\bibliographystyle{IEEEtran}
\bibliography{references}

@inproceedings{aidev,
  author    = {Hao Li and Haoxiang Zhang and Ahmed E. Hassan},
  title     = {{AIDev}: Studying {AI} Coding Agents on {GitHub}},
  booktitle = {Proceedings of the 23rd International Conference on Mining Software Repositories (MSR)},
  year      = {2026},
  doi       = {10.1145/3793302.3797249}
}

@misc{census,
  author = {Arsham Khosravani and Audris Mockus},
  title  = {Detecting {AI} Coding Agents in Open Source: A Validated Multi-Method Census of 180 Million Repositories},
  year   = {2026},
  eprint = {2606.24429},
  archivePrefix = {arXiv},
  note   = {arXiv:2606.24429}
}

@misc{replication,
  author       = {Anonymous},
  title        = {{Replication Package for ``Do AI Coding Agents Change Infrastructure Deployment Code Like Humans?''}},
  year         = {2026},
  howpublished = {\url{https://zenodo.org/records/23650457}}
}

@article{rahman2019mapping,
  author  = {Akond Rahman and Rezvan Mahdavi-Hezaveh and Laurie Williams},
  title   = {A Systematic Mapping Study of Infrastructure as Code Research},
  journal = {Information and Software Technology},
  volume  = {108},
  year    = {2019}
}

@article{ksontini2026docker,
  author  = {Ksontini, Emna and Ferreira, Thiago and
             Khalsi, Rania and Kessentini, Wael},
  title   = {Understanding Docker Refactorings:
             Expanded Taxonomy, Operational Trade-Offs,
             and Role-Aware Recommendations},
  journal = {IEEE Transactions on Software Engineering},
  volume  = {52},
  number  = {3},
  pages   = {786--808},
  year    = {2026},
  doi     = {10.1109/TSE.2025.3649731}
}

@inproceedings{rahman2019sins,
  author    = {Akond Rahman and Chris Parnin and Laurie Williams},
  title     = {The Seven Sins: Security Smells in Infrastructure as Code Scripts},
  booktitle = {Proceedings of the 41st International Conference on Software Engineering (ICSE)},
  year      = {2019}
}

@inproceedings{rahman2020gang,
  author    = {Akond Rahman and Effat Farhana and Chris Parnin and Laurie Williams},
  title     = {Gang of Eight: A Defect Taxonomy for Infrastructure as Code Scripts},
  booktitle = {Proceedings of the 42nd International Conference on Software Engineering (ICSE)},
  year      = {2020}
}

@inproceedings{sharma2016smells,
  author    = {Tushar Sharma and Marios Fragkoulis and Diomidis Spinellis},
  title     = {Does Your Configuration Code Smell?},
  booktitle = {Proceedings of the 13th International Conference on Mining Software Repositories (MSR)},
  year      = {2016}
}

@article{wessel2018bots,
  author  = {Mairieli Wessel and Bruno Mendes de Souza and Igor Steinmacher and Igor S. Wiese and Ivanilton Polato and Ana Paula Chaves and Marco A. Gerosa},
  title   = {The Power of Bots: Understanding Bots in {OSS} Projects},
  journal = {Proceedings of the ACM on Human-Computer Interaction},
  volume  = {2},
  number  = {CSCW},
  pages   = {182},
  year    = {2018},
  doi     = {10.1145/3274451}
}

@inproceedings{dey2020bots,
  author    = {Tapajit Dey and Sara Mousavi and Eduardo Ponce and Tanner Fry and Bogdan Vasilescu and Anna Filippova and Audris Mockus},
  title     = {Detecting and Characterizing Bots that Commit Code},
  booktitle = {Proceedings of the 17th International Conference on Mining Software Repositories (MSR)},
  year      = {2020}
}

@inproceedings{alfadel2021dependabot,
  author    = {Mahmoud Alfadel and Diego Elias Costa and Emad Shihab and Mouafak Mkhallalati},
  title     = {On the Use of Dependabot Security Pull Requests},
  booktitle = {Proceedings of the 18th International Conference on Mining Software Repositories (MSR)},
  year      = {2021}
}

@inproceedings{nguyen2022copilot,
  author    = {Nhan Nguyen and Sarah Nadi},
  title     = {An Empirical Evaluation of {GitHub} {Copilot}'s Code Suggestions},
  booktitle = {Proceedings of the 19th International Conference on Mining Software Repositories (MSR)},
  year      = {2022}
}

@inproceedings{pearce2022asleep,
  author    = {Hammond Pearce and Baleegh Ahmad and Benjamin Tan and Brendan Dolan-Gavitt and Ramesh Karri},
  title     = {Asleep at the Keyboard? Assessing the Security of {GitHub} {Copilot}'s Code Contributions},
  booktitle = {Proceedings of the IEEE Symposium on Security and Privacy (S\&P)},
  year      = {2022}
}

@inproceedings{saavedra2022glitch,
  author    = {Nuno Saavedra and Jo{\~a}o F. Ferreira},
  title     = {{GLITCH}: Automated Polyglot Security Smell Detection in Infrastructure as Code},
  booktitle = {Proceedings of the 37th IEEE/ACM International Conference on Automated Software Engineering (ASE)},
  year      = {2022}
}

@inproceedings{kosbar2025smellssus,
  author    = {Seif Kosbar and Mohammad Hamdaqa},
  title     = {Smells-sus: Sustainability Smells in {IaC}},
  booktitle = {Proceedings of the 22nd International Conference on Mining Software Repositories (MSR)},
  year      = {2025}
}

@article{verdet2025policies,
  author  = {Alexandre Verdet and Mohammad Hamdaqa and Leuson Da Silva and Foutse Khomh},
  title   = {Assessing the Adoption of Security Policies by Developers in {Terraform} across Different Cloud Providers},
  journal = {Empirical Software Engineering},
  volume  = {30},
  number  = {3},
  year    = {2025},
  doi     = {10.1007/s10664-024-10610-0}
}

@misc{war2025sins,
  author = {Aicha War and Serge L. B. Nikiema and Jordan Samhi and Jacques Klein and Tegawend{\'e} F. Bissyand{\'e}},
  title  = {Security Smells in Infrastructure as Code Scripts: A Taxonomy Update Beyond the Seven Sins},
  year   = {2025},
  note   = {arXiv:2509.18761}
}

@misc{rahman2018bugs,
  author = {Akond Rahman and Sarah Elder and Faysal Hossain Shezan and Vanessa Frost and Jonathan Stallings and Laurie Williams},
  title  = {Bugs in Infrastructure as Code},
  year   = {2018},
  note   = {arXiv:1809.07937}
}

@inproceedings{begoug2024jit,
  author    = {Mahi Begoug and Moataz Chouchen and Ali Ouni and Eman Abdullah AlOmar and Mohamed Wiem Mkaouer},
  title     = {Fine-Grained Just-In-Time Defect Prediction at the Block Level in Infrastructure-as-Code ({IaC})},
  booktitle = {Proceedings of the 21st International Conference on Mining Software Repositories (MSR)},
  year      = {2024}
}

@article{cliff1993delta,
  author  = {Norman Cliff},
  title   = {Dominance Statistics: Ordinal Analyses to Answer Ordinal Questions},
  journal = {Psychological Bulletin},
  volume  = {114},
  number  = {3},
  year    = {1993}
}

@inproceedings{romano2006cliff,
  author    = {Jeanine Romano and Jeffrey D. Kromrey and Jesse Coraggio and Jeff Skowronek},
  title     = {Appropriate Statistics for Ordinal Level Data: Should We Really Be Using t-test and {Cohen's} d for Evaluating Group Differences on the {NSSE} and Other Surveys?},
  booktitle = {Annual Meeting of the Florida Association of Institutional Research},
  year      = {2006}
}

@article{mcnemar1947,
  author  = {Quinn McNemar},
  title   = {Note on the Sampling Error of the Difference Between Correlated Proportions or Percentages},
  journal = {Psychometrika},
  volume  = {12},
  number  = {2},
  pages   = {153--157},
  year    = {1947}
}

@article{holm1979,
  author  = {Sture Holm},
  title   = {A Simple Sequentially Rejective Multiple Test Procedure},
  journal = {Scandinavian Journal of Statistics},
  volume  = {6},
  number  = {2},
  pages   = {65--70},
  year    = {1979}
}

@inproceedings{kon2024iaceval,
  author    = {Patrick Tser Jern Kon and Jiachen Liu and Yiming Qiu and Weijun Fan and Ting He and Lei Lin and Haoran Zhang and Owen M. Park and George S. Elengikal and Yuxin Kang and Ang Chen and Mosharaf Chowdhury and Myungjin Lee and Xinyu Wang},
  title     = {{IaC-Eval}: A Code Generation Benchmark for Cloud Infrastructure-as-Code Programs},
  booktitle = {Advances in Neural Information Processing Systems 37 (NeurIPS), Datasets and Benchmarks Track},
  year      = {2024}
}

@misc{nekrasov2025iacgen,
  author        = {Roman Nekrasov and Stefano Fossati and Indika Kumara and Damian Andrew Tamburri and Willem-Jan van den Heuvel},
  title         = {{IaC} Generation with {LLMs}: An Error Taxonomy and A Study on Configuration Knowledge Injection},
  year          = {2025},
  eprint        = {2512.14792},
  archivePrefix = {arXiv}
}

@inproceedings{vargas2026secureterraform,
  author    = {Francis Luis Santos Vargas and Rodrigo Brand{\~a}o Mansilha and Diego Kreutz},
  title     = {Can Language Models Generate Secure {Terraform} Code? {A} Security-Focused Benchmark Using Static Analysis},
  booktitle = {Proceedings of the Simp{\'o}sio de Infraestrutura Digital/Nuvem para Pesquisa (Pesquisa@Nuvem)},
  pages     = {29--38},
  year      = {2026},
  doi       = {10.5753/pesquisanuvem.2026.23954}
}

@misc{watanabe2025agentic,
  author        = {Miku Watanabe and Hao Li and Yutaro Kashiwa and Brittany Reid and Hajimu Iida and Ahmed E. Hassan},
  title         = {On the Use of Agentic Coding: An Empirical Study of Pull Requests on {GitHub}},
  year          = {2025},
  eprint        = {2509.14745},
  archivePrefix = {arXiv},
  note          = {arXiv:2509.14745}
}

@inproceedings{ehsani2026failed,
  author    = {Ramtin Ehsani and Sakshi Pathak and Shriya Rawal and Abdullah Al Mujahid and Mia Mohammad Imran and Preetha Chatterjee},
  title     = {Where Do {AI} Coding Agents Fail? {An} Empirical Study of Failed Agentic Pull Requests in {GitHub}},
  booktitle = {Proceedings of the 23rd International Conference on Mining Software Repositories (MSR)},
  pages     = {807--811},
  year      = {2026},
  doi       = {10.1145/3793302.3793579}
}

@misc{nakashima2026rejected,
  author        = {Sota Nakashima and Yuta Ishimoto and Masanari Kondo and Shane McIntosh and Yasutaka Kamei},
  title         = {Why Agentic-{PRs} Get Rejected: A Comparative Study of Coding Agents},
  year          = {2026},
  eprint        = {2602.04226},
  archivePrefix = {arXiv}
}

@article{wilson1927,
  author  = {Wilson, Edwin B.},
  title   = {Probable Inference, the Law of Succession, and Statistical Inference},
  journal = {Journal of the American Statistical Association},
  volume  = {22},
  number  = {158},
  pages   = {209--212},
  year    = {1927},
  doi     = {10.1080/01621459.1927.10502953}
}

@article{agresti2005matched,
  author  = {Agresti, Alan and Min, Yongyi},
  title   = {Simple Improved Confidence Intervals for Comparing Matched Proportions},
  journal = {Statistics in Medicine},
  volume  = {24},
  number  = {5},
  pages   = {729--740},
  year    = {2005},
  doi     = {10.1002/sim.1781}
}

@techreport{nist800204c,
  author      = {Ramaswamy Chandramouli},
  title       = {Implementation of {DevSecOps} for a Microservices-based Application with Service Mesh},
  institution = {National Institute of Standards and Technology},
  type        = {NIST Special Publication},
  number      = {800-204C},
  year        = {2022},
  doi         = {10.6028/NIST.SP.800-204C}
}

@article{pahl2026aiiac,
  author  = {Claus Pahl and {\"O}mer Can Sezen and Florian Hofer},
  title   = {Artificial Intelligence for Infrastructure-as-Code: A Systematic Literature Review},
  journal = {Electronics},
  volume  = {15},
  number  = {4},
  pages   = {755},
  year    = {2026},
  doi     = {10.3390/electronics15040755}
}

@article{esposito2026copilot,
  author  = {Matteo Esposito and Mikel Robredo and Alexander Bakhtin and Davide Taibi and Valentina Lenarduzzi},
  title   = {Generative {AI} as an Infrastructure Copilot: Automating Infrastructure-as-Code across the {DevSecOps} Lifecycle},
  journal = {Automated Software Engineering},
  volume  = {33},
  number  = {2},
  pages   = {58},
  year    = {2026},
  doi     = {10.1007/s10515-026-00600-5}
}

\end{document}